\documentclass[graybox,natbib,nosecnum]{svmult}
\bibpunct{(}{)}{;}{a}{}{,} 

\pdfoutput=1   

\usepackage{mathptmx}       
\usepackage{helvet}         
\usepackage{courier}        
\usepackage{type1cm}        

\usepackage{makeidx}         
\usepackage{graphicx}        
\usepackage{multicol}        
\usepackage[bottom]{footmisc}
\usepackage[normalem]{ulem}	
\usepackage{hyperref}  

\usepackage{soul}   

\newcommand{\hbindex}[1]{\hl{#1}\index{#1}}  

\makeindex             

\begin{document}

\title*{Planetary atmospheres Through Time: Effects of Mass Loss and Thermal Evolution}
\titlerunning{Planetary atmospheres Through Time} 
\author{Daria I. Kubyshkina}
\institute{Space Research Institute, Austrian Academy of Sciences, Schmiedlstrasse 6, A-8042 Graz, Austria, \email{ daria.kubyshkina@oeaw.ac.at}}
%
%
\maketitle
\abstract{\hbindex{Atmospheric mass loss} is a fundamental phenomenon shaping the structure and evolution of planetary atmospheres. It can engage processes ranging from global interactions with the host star and large-scale hydrodynamic outflows to essentially microphysical kinetic effects. The relevance of these processes is expected to change between planets of different properties and at different stages in planetary and stellar evolution. 
The early evolution of planetary atmospheres, as well as atmospheric escape from close-in planets hosting \hbindex{hydrogen-dominated atmospheres}, is thought to be driven by thermal hydrodynamic escape, while the kinetic non-thermal effects are most relevant for the long-term evolution of planets with secondary atmospheres, similar to the inner planets in the Solar System.
The relative input of different mechanisms, hence, the mass loss rate, shows a complicated dependence on planetary parameters and the parameters of the host star, where the latter evolve strongly with time. It results in a large variety of possible evolution paths of planetary atmospheres.}

\section{Introduction }
Within the last two decades, thanks to the exponentially increasing number and improving quality of observations {--} the results of the work of multiple missions dedicated to studying exoplanets and exoplanet-hosting stars, our understanding of planetary formation and evolution has greatly advanced. A range of theoretical models was developed to study the evolution of planetary atmospheres; at the same time, this development highlighted the lack of knowledge about different aspects of this process. 

The key process driving atmospheric evolution -- atmospheric mass loss -- can be caused by diverse processes of different nature and occur on {immensely} different timescales. Thus, the early evolution of primordial hydrogen-dominated atmospheres, associated with catastrophic losses and large-scale changes in observable planetary parameters (mass and radius), is thought to be driven by thermal \hbindex{hydrodynamic escape}, while the evolution of terrestrial-like secondary atmospheres (such as the atmospheres of Earth, Mars, and Venus) in Gyr-old systems is controlled by non-thermal \hbindex{ion escape} processes. Identifying the processes that contributed most to shaping the population of planets, as known to date, is a non-trivial task. It demands considering a wide range of factors affecting atmospheric mass loss, including the evolution of host stars, a variety of possible planetary compositions, and the planet's migration history.

This chapter briefly overviews the most common atmospheric escape processes and the factors controlling them. It proceeds by considering the common approach to modelling atmospheric evolution and the evolution paths of planets of specific types and concludes by discussing some possible imprints of atmospheric evolution onto the observed population of exoplanets.

 \section{Atmospheric escape processes}\label{sec:escape_processes}
Atmospheric mass loss occurs on different space- and timescales and can be driven by a wide range of physical processes, depending on the age of the planet and the type of its atmosphere. 
Formally, these processes can be divided into thermal and non-thermal ones. The former depend on the energy deposition in the atmosphere through heating, and the latter relate to microphysical interactions of the atmospheric species with each other, the magnetic field of planets, and stellar winds, or to the effects related to the tidal or magnetic interaction with host stars. 

Both thermal and non-thermal processes can take place at microphysical scales, considering the {loss} of individual particles energetic enough to escape the planet's gravity, or, as a bulk outflow, at atmospheric scales comparable to planetary radii, if the atmosphere remains collisional up to high altitudes, and hydrodynamic approach applies. Furthermore, the mass loss rates associated with different escape processes vary in a broad range. The mass loss due to non-thermal escape processes is typically by about an order of magnitude lower than due to thermal escape \citep[e.g.][]{Kislyakova2014A&A...562A.116K}; the thermal mass loss rates themselves, in turn, can vary by several orders of magnitude depending on the energy budget set by the parameters of the host star, the orbital distance of the planet, and its internal energy (e.g. the post-formation heat released from accretion and contraction of the core, or decay of the radioactive elements). Therefore, even though, given favourable conditions, different escape processes can occur in planetary atmospheres simultaneously, at a specific moment the atmospheric evolution is likely to be controlled by just one of them.

Recent studies of atmospheric evolution suggest that during the first few tens or hundreds of megayears, the escape of primordial hydrogen-dominated atmospheres from close-in planets \citep[planets with orbital periods shorter than $\sim100$ days, dominating the current observational sample; e.g.][]{Fulton2017AJ....154..109F,Fulton2018AJ....156..264F,Berger2018ApJ...866...99B,Stassun2019AJ....158..138S,Petigura2022AJ....163..179P} occurs in form of hydrodynamic outflow fuelled by the heating from {stellar high-energy radiation} (X-ray and extreme ultraviolet, EUV, together, XUV) or by the own thermal energy of the planet. Following the relatively short phase of extreme hydrodynamic escape, atmospheric evolution can take different paths depending on the planet type. A large fraction of low-mass planets (less massive than $\sim10$\,$M_{\oplus}$) can have lost their primordial atmospheres already at this point, after which the \hbindex{secondary atmospheres} can be built through the outgassing from planetary interiors -- to be further removed through the thermal escape processes or to be preserved for the lifetime of the system, depending on the atmospheric composition and conditions at the planet. The evolution of close-in low- to intermediate-mass planets ($<\sim100$\,$M_{\oplus}$) that can preserve some of their primary atmospheres, can be dominated by hydrodynamic \hbindex{XUV-driven escape} on gygayears timescales \citep[e.g.][]{Rogers2021MNRAS.508.5886R,Owen_Schlichting2023MNRAS.tmp.3804O}, while the escape from cooler and/or heavier planets, as well as that in the evolved planetary systems, can be dominated by kinetic thermal or non-thermal escape processes. The following sections give a brief overview of different escape processes, and the reader is encouraged to study the recent reviews by \citet{Gronoff2020JGRA..12527639G}, \citet{Owen2020SSRv..216..129O}, and \citet{Koskinen2022ApJ...929...52K} for more details.

\subsection{Thermal escape}\label{sec:thermal_escape}
The longest-known form of thermal escape is the kinetic \hbindex{Jeans escape} \citep[e.g. ][]{Oepik1963GeoJ....7..490O,Chamberlain1963P&SS...11..901C,Mihalas_Mihalas1984oup..book.....M}. This approach assumes that the atmosphere can be described by Boltzmann distribution, and the particle escape occurs from the region around the exobase, the altitude level where the atmosphere becomes collisionless (i.e. the mean free path of a particle equals the pressure scale height). It also relies on the assumption that only a small fraction of atmospheric particles have energies sufficient to overcome the escape velocity, and the whole atmosphere undergoes no significant bulk motion. These assumptions hold for compact, terrestrial-like, secondary atmospheres or hydrogen-dominated atmospheres of Jupiter-like planets in the absence of significant energy input from an external source. 

The accurate treatment of this problem implies solving the Boltzmann equation \citep[see, e.g.][]{Volkov2011ApJ...729L..24V}; however, given the computational costs of this method, a common simplification considers the atmosphere as Maxwellian and isothermal \citep{Mihalas_Mihalas1984oup..book.....M}
\begin{equation}
    f(\vec{x}, \vec{v}) = n\left( \frac{1}{\pi u_{\rm th}^{2}}\right) ^{3/2}\exp\left( -\frac{v^2}{v_{\rm th}^2} \right) \,,
\end{equation}
where $n$ is the numerical density and $v_{\rm th} = \sqrt{\frac{2k_{\rm b}T}{\mu}}$ is the {thermal velocity} of the atmospheric species with the mass $\mu$. The particles escape the atmosphere if their velocities $v > v_{\rm esc} = \sqrt{\frac{2GM_{\rm pl}}{r}}$ at the \hbindex{exobase} (where $M_{\rm pl}$ is the planetary mass and $r$ is the radial distance from the planet centre). Thus, the flux of escaping particles depends on the relation between the {escape velocity} and the thermal speed
\begin{eqnarray}
    \Phi = n\left( \frac{v_{\rm th}^{2}}{4\pi}\right)^{1/2}(1 + \lambda_{\rm exo})\exp(-\lambda_{\rm exo})\,,\label{eq:phi_jeans}\\
    \lambda_{\rm exo} = \frac{v_{\rm esc}^2}{v_{\rm th}^{2}} = \frac{GM_{\rm pl}\mu}{k_{\rm b}T_{\rm exo}r_{\rm exo}}\,.\label{eq:lambda_exo}
\end{eqnarray}
The so-called \hbindex{Jeans parameter} $\lambda_{\rm exo}$ represents, in a more general sense, the relation of the planet's gravitational energy to the thermal energy deposited in its atmosphere, calculated at exobase. In exoplanetary studies, it is common to generalise this parameter to the photosphere (i.e. to substitute $r_{\rm exo}$  with planetary radius $R_{\rm pl}$ in Equation\,\ref{eq:lambda_exo}; this parameter is further referred to as $\Lambda$). For compact terrestrial-like atmospheres, these two parameters are nearly equivalent; however, for close-in sub-Neptune-like planets and hot Jupiters, $r_{\rm exo}$ and $R_{\rm pl}$ (hence, $\lambda_{\rm exo}$ and $\Lambda$) can differ by the factor of a few or a few tens.
Both parameters are commonly used to get a coarse guess on the atmospheric mass loss rate (associated with thermal escape) and to distinguish between the Jeans and the hydrodynamic regimes. The border value of $\lambda_{\rm exo}$ was estimated 1-3 upon different assumptions \citep[e.g.][]{Volkov2011ApJ...729L..24V,Erkaev2015MNRAS.448.1916E}.

The formulation of the Jeans escape presented above has a range of limitations. The real distribution of the atmospheric particles is not quite Maxwellian, as the fastest particles are constantly removed from the atmosphere, and the atmosphere is not isothermal; escape also does not occur only from exobase. The range of corrections exists to account for these facts \citep[see ][ and the references therein]{Gronoff2020JGRA..12527639G} but in the most general case, one has to solve the Boltzmann equation to get the most accurate estimate. 

With increasing energy input, the thermal energy of atmospheric gas grows. 
When the mean energy of atmospheric particles becomes high enough to overcome both the gravity of the planet and the binding due to the collisions between themselves, the atmospheric escape can engage deep below the exobase level, taking the form of a continuous bulk outflow \citep[e.g. ][]{watson1981Icar...48..150W}. 
Thus, the escape transitions from Jeans-like to hydrodynamic and can be treated assuming collisional, fluid-like atmospheres, as considered in more detail in Section\,`Hydrodynamic escape'.

Differently to the hydrodynamic escape, where the dense outflow of light elements (such as hydrogen and helium) can drag away the heavier elements \citep[e.g. ][]{Zahnle_Kasting1986Icar...68..462Z,Pepin1991Icar...92....2P,Odert2018Icar..307..327O,Lammer2020Icar..33913551L}, in ``collisionless'' Jeans escape the escape of different species occurs almost independently. Therefore, in quiet conditions, Jeans escape can describe the escape of lighter particles through the stable heavy-element atmospheres, such as the escape of hydrogen and helium from the present-day Earth, Venus, and Mars \citep{Shizgal1986P&SS...34..279S,Chamberlain1969ApJ...155..711C,Hunten1973JAtS...30.1481H}. This makes Jeans escape particularly important for the fractionation of light elements in planetary atmospheres.

\subsection{Non-thermal escape}\label{sec:non-thermal_escape}

Non-thermal escape processes can be formally split into three groups: the tidal and magnetic interaction with the host star, interaction with stellar winds, and the ion escape processes at magnetized and non-magnetized planets. 
The tidal interactions with the host star become relevant for atmospheric escape from planets in very short orbits (within $\sim0.1$\,AU for a typical sub-Neptune-like planet around FGKM stars), and their influence can be twofold. First, stellar gravity forces affect the atmospheric material directly, decreasing the energy needed for atmospheric particles to leave the gravitation well of the planet \citep[see, e.g.][]{Erkaev2007A&A...472..329E}. In extreme cases of very small orbital distances, the planetary Roche lobe (wherein the gravity of the planet dominates over that of its host star) can reside in relatively low atmospheric layers with pressures of $\sim10-100$\,nbar, resulting in atmospheric disruption through the \hbindex{Roche lobe overflow} \citep[e.g.][]{Koskinen2022ApJ...929...52K,Huang2023ApJ...951..123H}. Second, the {tidal heating} in deep atmospheric layers \citep[e.g.][]{Bodenheimer2001ApJ...548..466B,Arras2010ApJ...714....1A} can lead to planetary inflation and intensification of thermal and non-thermal escape processes due to the increase of the energy budget of the atmosphere itself and the increase of the interaction surface with stellar high-energy photons and stellar wind.
A similar effect can be caused by the magnetic interaction between the star and a close-in planet, through the induction of currents in the deep atmospheric layers \citep[e.g. ][]{Batygin2010ApJ...714L.238B,Wu_Lithwick2013ApJ...763...13W,Ginzburg2016ApJ...819..116G,Kislyakova2018ApJ...858..105K}. For further detail on these processes, the reader is advised to check the recent review by \citet{Fortney2021JGRE..12606629F}.%

\hbindex{Stellar winds} represent the hot coronal material (i.e. low-density
plasma) accelerated to high velocities and carrying stellar magnetic field \citep[see, e.g. ][]{Vidotto2015MNRAS.449.4117V}. Their interaction with planetary atmospheres depends strongly on the presence of the planetary magnetic field, the mass, radius, and orbital parameters of the planet, and the age of the system \citep[e.g. ][]{Schneiter2007ApJ...671L..57S,Khodachenko2012ApJ...744...70K,Khodachenko2015ApJ...813...50K,Khodachenko2019ApJ...885...67K,Shaikhislamov2014ApJ...795..132S,Carolina2014MNRAS.438.1654V,Carolina2018MNRAS.479.3115V,Carolina2021MNRAS.501.4383V,Tripathi2015ApJ...808..173T,Carroll-Nellenback2017MNRAS.466.2458C,Vidotto2018MNRAS.481.5296V,Vidotto2021LRSP...18....3V,Daley-Yates2019MNRAS.483.2600D,Debrecht2019MNRAS.483.1481D,Esquivel2019MNRAS.487.5788E,McCann2019ApJ...873...89M,Carolan2020MNRAS.498L..53C,Cohen2022ApJ...934..189C,Kubyshkina2022MNRAS.510.2111K}. Along with the erosion of planetary atmospheres, extra heating of atmospheres by stellar wind energetic particles, and \hbindex{ion pick-up} \citep[drag of ions from the exosphere due to the interaction with the electromagnetic field of the wind; e.g.][]{Kislyakova2014A&A...562A.116K}, increasing the atmospheric mass loss, stellar winds can confine planetary atmospheres leading to the reduction of the escape if the pressure balance is reached close to the planet \citep[e.g. ][]{Christie2016ApJ...820....3C,Vidotto2020MNRAS.494.2417V,Carolan2020MNRAS.498L..53C,Carolan2021MNRAS.500.3382C}.

The range of non-thermal escape processes represents the escape of ionised atmospheric species, which typically dominate planetary exospheres (mainly due to photoionisation by stellar photons). These processes depend crucially on the strength and configuration of the \hbindex{planetary magnetic field}, as ions can be trapped and accelerated on the closed magnetic field lines \citep[e.g.][]{Matsakos2015A&A...578A...6M,Carolina2018MNRAS.479.3115V,Cohen2022ApJ...934..189C}.
Thus, the presence of magnetic fields affects the precipitation of energetic particles and ions from upper atmospheres, interaction with energised particles from stellar winds and/or the ionosphere, processes driven by photochemistry, and the overall dynamics of ionised material in the vicinity of the planet. 
However, the total effect of the magnetic field on atmospheric mass loss is hard to quantify due to the large number of processes involved. 
For terrestrial planets in the Solar System (dominated by non-thermal escape processes), it was shown that a weak magnetic field can intensify the escape, while the strong field is, in general, expected to be protective \citep{Gunell2018A&A...614L...3G,Sakai2018GeoRL..45.9336S,Egan2019MNRAS.488.2108E,Ramstad2021SSRv..217...36R}. 
For Hot Jupiters, where planetary winds are driven by hydrodynamic processes, models predict significant suppression of the
atmospheric mass losses for magnetic field strength above 0.1--3 G \citep[see ][]{Owen_Adams2014MNRAS.444.3761O,Trammell2014ApJ...788..161T,Khodachenko2015ApJ...813...50K,Khodachenko2021MNRAS.507.3626K,Arakcheev2017ARep...61..932A}. 

In the absence of a magnetic field, the escape of ions occurs similarly to the Jeans-like escape, meaning that only the ions having sufficient energies can leave the planet. The mechanisms of the energising of ions can be represented by photochemical reactions and collisions with energetic stellar wind particles \citep[e.g. ][]{Shizgal1986P&SS...34..279S,Shematovich1994JGR....9923217S,Johnson1994SSRv...69..215J,Johnson2008SSRv..139..355J,Lee2015JGRE..120.1880L}. Though the atmospheric mass loss rates associated with non-thermal escape processes are low compared to the thermal ones, and these processes are, in general, not expected to be capable of removing the atmosphere from a planet and therefore considered irrelevant for the long-term evolution of primordial hydrogen-dominated atmospheres, they are important for the formation of secondary atmospheres \citep[see, e.g.][ for further details]{Gronoff2020JGRA..12527639G}. 

\section{Hydrodynamic escape}\label{sec:hydro_escape}
In contrast with Jeans-like escape, the hydrodynamic approach treats an atmosphere as a collisional, fluid-like medium. 
For decades, hydrodynamics were applied to study the early evolution of Earth and Venus \citep[see, e.g.][]{Dayhoff1967Sci...155..556D,Sekiya1980PThPh..64.1968S,watson1981Icar...48..150W,Tian2005Sci...308.1014T}, and later on this approach was generalised on the hydrogen-dominated atmospheres of \hbindex{sub-Neptune-like} and giant \hbindex{extrasolar planets} \citep[see, e.g.][]{Lammer2003ApJ...598L.121L,Lecavelier2004A&A...418L...1L,Baraffe2004A&A...419L..13B,Yelle2004Icar..170..167Y,Erkaev2007A&A...472..329E,Erkaev2015MNRAS.448.1916E,Erkaev2016MNRAS.460.1300E,GMunoz2007P&SS...55.1426G,GMunoz2019ApJ...884L..43G,GMunoz2023A&A...672A..77G,Penz2008A&A...477..309P,Cecchi-Pestellini2009A&A...496..863C,Murray-Clay2009ApJ...693...23M,Ehrenreich2011A&A...529A.136E,Owen_Jackson2012MNRAS.425.2931O,kubyshkina2018A&A...619A.151K,Caldiroli2021A&A...655A..30C,Schulik_Booth2023MNRAS.523..286S}. 
For substantial hydrogen-dominated atmospheres of close-in planets, the hydrodynamic escape represents the major source of the atmospheric mass loss, overcoming by orders of magnitude kinetic and non-thermal escape processes. Therefore, it is considered the main driving mechanism of the {atmospheric evolution} \citep[e.g.][]{Murray-Clay2009ApJ...693...23M,Lopez2012ApJ...761...59L,Chen2016ApJ...831..180C,kubyshkina2019A&A...632A..65K,kubyshkina2019ApJ...879...26K,kubyshkina2022MNRAS.510.3039K,Pezzotti2021A&A...654L...5P,Gu_Chen2023ApJ...953L..27G} and plays a major role in shaping the population of low to intermediate mass exoplanets as it is known to date \citep[e.g.][]{Fulton2017AJ....154..109F,Fulton2018AJ....156..264F,Owen_Wu2017ApJ...847...29O,gupta_schlichting2019MNRAS.487...24G,Mordasini2020A&A...638A..52M}.

The hydrodynamically escaping atmosphere can be described with a set of fluid dynamics equations of the mass, momentum, and energy conservation as
\begin{eqnarray}
\frac{\partial\rho}{\partial t} + \frac{\partial(\rho v r^2)}{r^2\partial r} &=& 0\,, \label{eq:mass_cons}\\
\frac{\partial\rho v}{\partial t} + \frac{\partial[r^2(\rho v^2+P)]}{r^2\partial r} &=& - \frac{\partial U}{\partial r} + \frac{2P}{r}\,, \label{eq:mom_cons}\\
\frac{\partial E}{\partial t} + \frac{\partial[vr^2(E + P)]}{r^2\partial r} &=& Q  + \frac{\partial}{r^2\partial r}(r^2\chi \frac{\partial T}{\partial r}) - \frac{\partial(\rho U)}{r^2\partial r}\,.\label{eq:en_cons}
\end{eqnarray}
The variables are the radial distance from the planet centre ($r$), density ($\rho$), temperature ($T$), and bulk velocity ($v$) of the atmosphere, thermal pressure ($P$), the sum of the kinetic and thermal energies ($E$), and the gravitational potential $U$. The latter accounts for the gravitational potential of the central body (planet) and the gravitational forces of the host star \citep[relevant for close-in planets][]{Erkaev2007A&A...472..329E}. 
The second term on the right-hand of Equation\,\ref{eq:en_cons} accounts for the thermal conductivity of the neutral gas.%

Parameter $Q$ in Equation\,\ref{eq:en_cons} represents the sum of all heating and cooling rates given by (chemical) processes accounted for in the specific model. 
For hydrogen-dominated atmospheres, the main heating source is typically considered the \hbindex{photoionisation} of hydrogen atoms by stellar photons
\begin{equation}
    {\mathrm H} + h\nu \rightarrow {\mathrm H^+ + e^*}\,\label{eq:photoion}.
\end{equation}
If the energy of a photon exceeds the ionisation threshold of hydrogen (which is the case for XUV photons), it leads to the production of an ion $H^+$ and a free electron, that carries away the excess energy $h\nu-13.6$\,eV. This excess energy can be further lost in collisions (heating the atmosphere) or spent on secondary ionisation and excitation of atmospheric species. To accurately define which fraction of the absorbed photon energy is spent on heating, one has to account for all relevant reactions and also consider all (or a sufficient number of) energy levels of the atmospheric species, as the energy of an electron tore away in reaction\,\ref{eq:photoion} can vary in a wide range. The resulting \hbindex{heating efficiency}, i.e. the relation of the energy spent on heating to the total absorbed energy, varies with altitude and the wavelength (energy) of the incident radiation, and also depends on the type of a planet and atmospheric content \citep[e.g.][]{Dalgarno1999ApJS..125..237D,Yelle2004Icar..170..167Y,Shematovich2014A&A...571A..94S,Ionov2015SoSyR..49..339I,Salz2015A&A...576A..21S,Salz2016A&A...585L...2S}. Given the complexity and the numerical costs of this approach, the estimation of heating efficiency is often reduced to a constant coefficient $\eta$  \citep[e.g.][]{Owen_Jackson2012MNRAS.425.2931O} or an analytical dependence on the photon wavelength \citep[e.g.][]{Murray-Clay2009ApJ...693...23M}. In this case, the volume heating rate by the stellar flux of wavelength ${\mathrm\nu}$ can be written as
\begin{equation}\label{eq:photoheating_function}
    H_{\mathrm ion} = \frac{\eta\sigma_{\mathrm\nu}n}{2}\int_{0}^{\frac{\pi}{2}+\arccos(1/r)}I_{\mathrm\nu}(r,\theta)\sin\theta d\theta\,,
\end{equation}
where $\sigma_{\mathrm\nu}$ is the wavelength-dependent absorption cross-section of hydrogen atoms, and $I_{\mathrm\nu}(r,\theta)$ describes the spatial variations of the stellar irradiation due to the atmospheric absorption in spherical coordinates.

Other heating and cooling sources are given by other exothermic and endothermic reactions (e.g. \hbindex{Ly$\alpha$-cooling} ${\mathrm H + e^* \rightarrow H^* + e}$),
where the corresponding cooling/heating rates are given by the product of the numerical densities of participating species and the specific reaction rates. %
To account for these processes (which typically include photodissociation, recombination, collisional ionisation, etc.), in parallel with Equations\,\ref{eq:mass_cons}--\ref{eq:en_cons} one has to solve the equations describing chemical equilibrium. They take the form of Equation\,\ref{eq:mass_cons} and account for the sink and replenishment of each atmospheric species according to chemical reactions.

Until recently, the most common and widely studied XUV-driven hydrodynamic escape, assuming the photoionisation heating given by Equation\,\ref{eq:photoheating_function} to solely power the hydrodynamic wind, was considered the main driver of atmospheric evolution \citep[e.g.][]{Lopez2012ApJ...761...59L,Lopez_Fortney2013ApJ...776....2L,Chen2016ApJ...831..180C,Owen_Wu2017ApJ...847...29O}.
In this approach, the heating by stellar XUV is expected to occur in the narrow altitude range close to the planetary photosphere and lead to the adiabatic expansion and settling of the transonic planetary wind. In this case, it is common to reduce this range to one specific height $R_{\rm eff}$ and approximate the atmospheric mass loss rates using the \hbindex{energy-limited approximation} \citep[e.g. using the form of][]{Erkaev2007A&A...472..329E}
\begin{equation}\label{eq:EL}
    \dot{M}_{\rm EL} = \pi\frac{\eta F_{\rm XUV}R_{\rm eff}^{2}R_{\rm pl}}{KGM_{\rm pl}}\,,
\end{equation}
where parameter $K\leq 1$ accounts for stellar gravity effects. This approximation provides a good enough estimate for the atmospheric mass loss rates from highly irradiated massive planets. However, in the case of low-mass (low-density) planets, the absorption region of stellar XUV can be neither narrow nor close to the photosphere. For these planets, applying Equation\,\ref{eq:EL} (in particular, under common simplification $R_{\rm eff}\simeq R_{\rm pl}$) can lead to the underestimation of the mass loss rates by up to an order of magnitude \citep[for details see e.g.][]{Krenn2021A&A...650A..94K}.

Furthermore, along with stellar XUV heating, the internal thermal energy of the planet (acquired through the accretion of solids and contraction during the formation stage or from the stellar bolometric heating) can contribute largely to fuelling the hydrodynamic outflow. To account for this type of atmospheric escape, \citet{Ginzburg2016ApJ...819..116G,Ginzburg2018MNRAS.476..759G} introduced the term \hbindex{``core-powered mass loss''} and suggested the post-formation luminosity as the heating source. \citet{gupta_schlichting2019MNRAS.487...24G} generalised this idea for the bolometric heating from the host star and introduced an analytical approximation to estimate the mass loss rates using the basic planetary parameters (mass, radius, and the equilibrium temperature). The latter formulation is similar to that of the \hbindex{``boil-off''} atmospheric evaporation suggested by \citet{Owen_Wu2016ApJ...817..107O}; following \citet{Owen_Schlichting2023MNRAS.tmp.3804O}, both types of the escape driven by the energy deposited in planetary core or lower atmosphere are further referred to as the core-powered mass loss. 
For hot and inflated low-mass planets, the core-powered mass loss rates can overcome the XUV-driven escape (and predictions of Equation\,\ref{eq:EL}) by orders of magnitude. However, given that both the post-formation luminosity and the inflation degree of a planet decline rapidly with time, the input from this atmospheric escape mechanism relative to the XUV-driven escape \citep[that occurs on Gyr timescale, see e.g.][]{King_Wheatley2021MNRAS.501L..28K} is expected to maximise for planets younger than hundred Myr. 
Hydrodynamic simulations show that the transition from core-powered to XUV-driven mass loss mechanism occurs when the gravitational parameter $\Lambda$ drops below 10--30 \citep[e.g.][]{kubyshkina2018A&A...619A.151K,Owen_Schlichting2023MNRAS.tmp.3804O}.

\begin{figure}
    \centering
    \includegraphics[width=1\linewidth]{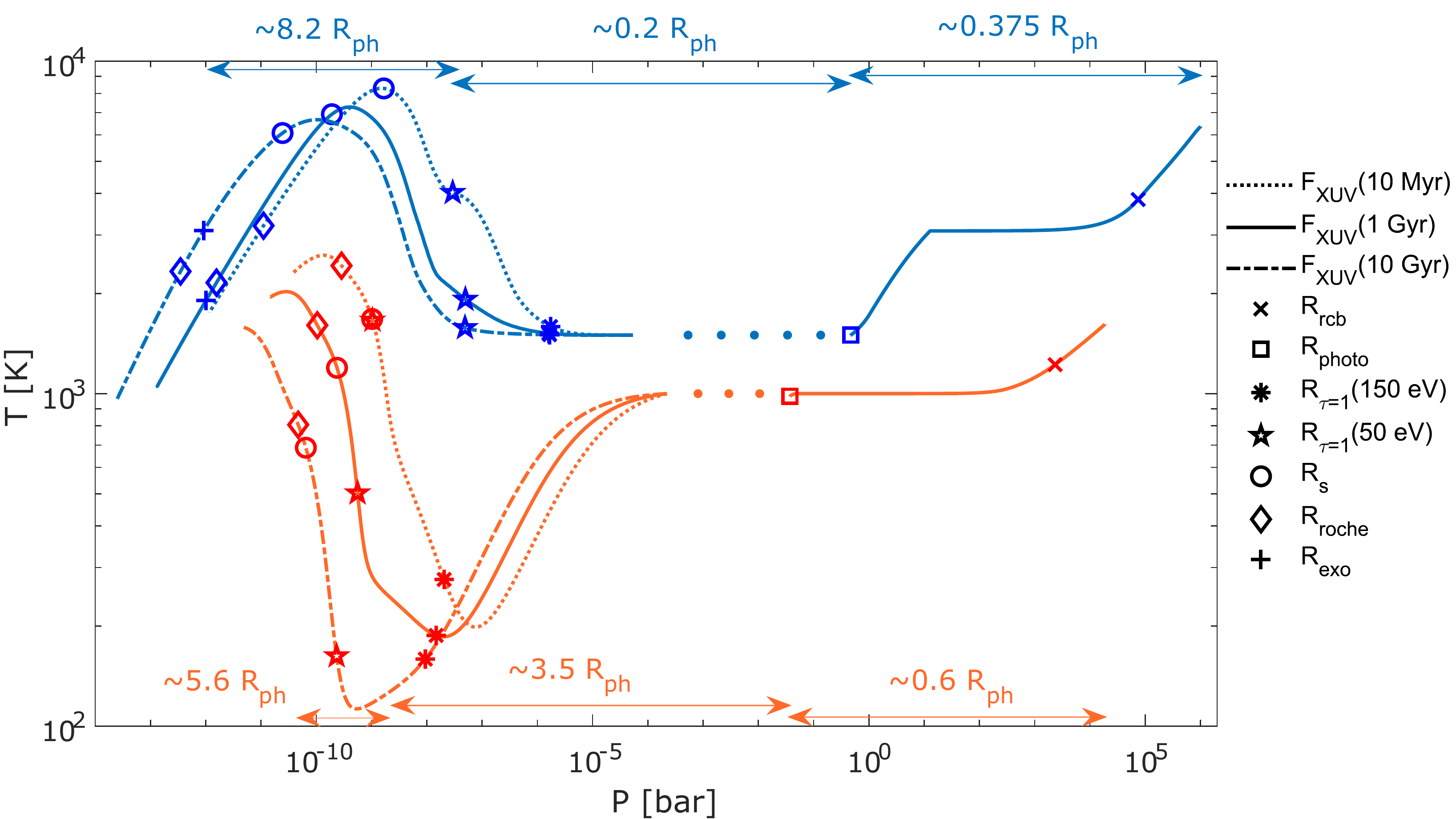}
    \caption{Pressure-temperature profiles of the sub-Saturn-like planet in the $\sim0.05$\,AU orbit (blue lines and symbols) and the low-mass super-puff planet in the $\sim0.1$\,AU orbit (orange lines and symbols). In both cases, the planet orbits the Sun-like host star. The profiles in the upper atmosphere ($P\leq10^{-1}$\,bar) are shown assuming the planet to be exposed to stellar XUV irradiation typical for the ages of 10\,Myr, 1\,Gyr, and 10\,Gyr, as indicated in the legend. The symbols denote the characteristic distances of the atmospheres, including radiative-convective boundary ($R_{\rm rcb}$), photosphere ($R_{\rm photo}$), absorption depths of 150\,eV  and 50\,eV photons ($R_{\rm tau=1}$), sonic point ($R_{\rm s}$), Roche radius ($R_{\rm roche}$), and the exobase position ($R_{\rm exo}$) where applicable. The arrows of the respective colours indicate the spatial scale of different atmospheric regions, from the right to the left: lower atmosphere, between the photosphere and the base of the thermosphere, and between the base of the thermosphere and the Roche radius.}
    \label{fig:PT}
\end{figure}
Figure\,\ref{fig:PT} shows the atmospheric pressure-temperature profiles for two very different planets: the massive {sub-Saturn-like planet} ($R_{\rm pl} = 4.0\,R_{\oplus}$ and $M_{\rm pl} = 45.1\,M_{\oplus}$, with $\sim20$\% of its mass in the atmosphere) and the low-mass {super-puff} ($R_{\rm pl} = 3.0\,R_{\oplus}$ and $M_{\rm pl} = 2.1\,M_{\oplus}$, with $\sim1$\% of its mass in the atmosphere). In the latter case, the atmosphere of the planet is strongly inflated due to the high luminosity of the core (presumably, post-formation luminosity), while for the sub-Saturn-like planet, any extra heating appears insufficient to counteract the planetary strong gravity. These profiles are plot employing the updated hydrodynamic models by \citet{kubyshkina2018A&A...619A.151K} including the detailed spectral energy distribution for the upper atmospheres ($P<\sim0.1$\,bar), and the MESA \citep[Modules for Experiments in Stellar Astrophysics;][]{Paxton2018ApJS..234...34P} for the lower atmospheres ($P>\sim0.1$\,bar). For upper atmospheres, the three different XUV irradiation levels are considered, as typical for a moderate solar-mass star at the ages of 10\,Myr, 1\,Gyr, and 10\,Gyr (shown by different line types). %

To illustrate the atmospheric structure and its relation to the atmospheric escape, Figure\,\ref{fig:PT} includes a range of characteristic distances and radial scales. Thus, the lower atmosphere ($P\geq\sim0.1$\,bar) is characterised by its extension and the parameters near the \hbindex{radiative-convective boundary} ($R_{\rm rcb}$), which essentially define the parameters of the core-powered mass loss. In both cases shown in Figure\,\ref{fig:PT}, the extension of the lower atmosphere is about $1.5\,R_{\oplus}$, though their masses are about a factor of 200 different. The position of the \hbindex{photosphere} ($R_{\rm ph}\simeq R_{\rm pl}$) varies between a few tens (for low-mass hot planets) to a few hundreds (for massive and/or cool planets) of millibar. 

For high-gravity planets, the region above the photosphere remains nearly isothermal, while the pressure and density decline steeply with the radial distance. The conditions in this region are favourable for condensation \citep[hence, formation of clouds/aerosols, see, e.g.][]{Koskinen2022ApJ...929...52K}, and most of the atmospheric species are expected to be in molecular form. For the sub-Saturn-like planet shown in Figure\,\ref{fig:PT}, the pressure changes by about 5 orders of magnitude within $\sim0.2R_{\rm ph}$. Due to these steep gradients, the atmosphere becomes transparent to high-energy stellar radiation close to the photosphere, and the XUV heating occurs within the narrow altitude range. Figure\,\ref{fig:PT} highlights the absorption heights $R_{\rm \tau=1}$ of the photons with energies of 150\,eV (roughly the boundary between the X-ray and EUV ranges) and 50\,eV (roughly the lower boundary of high-energy EUV). Thus, the X-ray radiation is absorbed below the former point, while the soft EUV radiation is absorbed above the latter. One can see, that the heating from the X-ray part of the spectrum appears negligible for all levels of stellar XUV, while the high-energy EUV provides a considerable input only for the highest XUV levels typical for young stars. Most of the heating, in turn, is provided by soft EUV photons. In all cases, the growth of the temperature proceeds up to the point where the speed of accelerating outflow reaches the sonic velocity (which is, $\sim10$\,km/s for the given case). Above this point, the adiabatic cooling due to the expansion of the atmosphere overtakes the XUV heating and the atmosphere cools down. The last two characteristic points that are included in Figure\,\ref{fig:PT} can be considered as the natural upper boundaries for the hydrodynamic outflow. The first one is the \hbindex{Roche radius} ($R_{\rm roche}$) of the planet where the gravitational forces from the planet and its host star equalise along the star-planet line. Any particle that trespassed on this boundary does not belong to the planet anymore. The second one is the position of the \hbindex{exobase} ($R_{\rm exo}$), which can be only seen in Figure\,\ref{fig:PT} for the high-gravity planet under XUV-irradiation typical for Gyr-old stars. At this point the atmosphere becomes collisionless, and, technically, the hydrodynamic approach is not valid anymore. In practice, the outflow still occurs hydrodynamically if this point is located above the \hbindex{sonic point} $R_{\rm s}$, as a signal can not propagate backwards in the \hbindex{supersonic outflow}.

In the case of a low-mass super-puff planet, the region above the photosphere is not isothermal. Due to the low gravity and high thermal energy of the planet, the atmosphere is dominated by adiabatic expansion, which leads to the strong cooling of the atmosphere and much more shallow pressure and density profiles. Thus, the density falls sufficiently to allow for the penetration of stellar high-energy photons over $\sim3.5\,R_{\rm ph}$, and the XUV absorption occurs at high altitudes and over a wide interval of radial distances. For this reason, the input from the high-energy EUV photons (50--150\,eV) becomes significant, while the input from X-ray remains negligible. It also limits the effect of soft EUV radiation, as it can only penetrate the uppermost atmospheric layers within the Roche radius. However, the model employed here assumes the constant heating efficiency for the different photon energies and altitudes, while more sophisticated models suggest that the heating efficiency is reduced for the photons with energies larger than 50\,eV because the high-energy photoelectrons are more likely to cause the secondary ionisation/excitation of atmospheric species rather than heating \citep[see, e.g.][]{GMunoz2023A&A...672A..77G}. Finally, as the outflow from this low-mass inflated planet is slower and denser compared to the sub-Saturn-like planet (density and velocity at the Roche radius are $\sim10^8$\,g/cm$^3$ and $\sim3-8$\,km/s against $10^5-10^7$\,g/cm$^3$ and $\sim20$\,km/s for the massive planet), the sonic point is located near the Roche radius. If the mean density of the planets decreases further and/or the equilibrium temperature of the planet increases, the strength of the core-powered outflow controlling the lower part of its upper atmosphere intensifies. This leads to a further increase in density at high altitudes and, at some point, prohibits the XUV radiation from penetrating the planetary upper atmosphere. Then the planet transitions to pure core-powered atmospheric escape.

These two fictional planets illustrate well the XUV-driven hydrodynamic mass loss (sub-Saturn-like planet) and the transition regime between the core-powered and XUV-driven mass loss (super-puff). In the first case, the atmospheric mass loss rates (ranging from $\sim9\times10^9-2.8\times10^{11}$\,g/s) correlate strongly with the XUV flux exposed to the planet. For both planets, the total XUV flux changes by the factor of 30 between the 10\,Myr and 10\,Gyr case; due to the relation between the EUV and X-ray fluxes changing with time \citep[e.g.][]{Sanz-Forcada2011A&A...532A...6S,Johnstone2021A&A...649A..96J}, this corresponds to the change in X-ray by the factor of 100, and the change in EUV by the factor of 20. Over the Gyr ages, the correlation between the EUV flux and the escape rates is approximately linear ($\dot{M}\sim F_{\rm EUV}$), while the input from X-ray remains negligible. For earlier ages, however, the whole XUV range has to be accounted for (though the X-ray effect remains secondary).
For the super-puff planet, the shift of the absorption to the higher altitudes increases the interaction surface with stellar high-energy photons ($\sim R_{\rm \tau=1}^2(E_{\rm\nu})$), which is expected to enhance the XUV heating \citep[e.g.][]{Owen_Schlichting2023MNRAS.tmp.3804O}; however, in the considered case the penetration of soft EUV photons is limited, hence, the input from XUV remains lower or comparable to the core-powered mass loss, and the dependence of the mass loss rates (ranging from $\sim4.2\times10^{11}-2.4\times10^{12}$\,g/s) on XUV/EUV flux is weaker than linear.

\section{Impact of the host star's evolution}\label{sec:stellar_evo}
Summarising the discussion above, the atmospheric evolution of close-in planets (in particular, of those with low to intermediate masses) is largely controlled by hydrodynamic escape, which, in turn, depends strongly on stellar XUV on gygayear timescales. Therefore, the evolution of planetary atmospheres is closely connected with the evolution of their \hbindex{host stars} or, more specifically, the evolution of stellar high-energy emission (\hbindex{chromospheric activity}). The latter, despite following the common trend for potentially planet-hosting FGKM stars (stellar XUV emission declining with time), can vary significantly between individual stars even if they have similar masses. 

Observations in young open clusters indicate that for the first tens or hundreds of megayears, stars remain in the saturation regime characterised by a weak dependence of the XUV luminosity ($L_{\rm XUV}$) on time and stellar rotation rate. As the star ages and its rotation slows down, reaching some critical threshold, the star leaves this regime and the relation of stellar X-ray to bolometric luminosities ($\frac{L_{\rm X}}{L_{\rm bol}}$) decays as a power law with age $\sim age^{-\alpha}$, where $\alpha$ is a constant ranging between $\sim1.1-2.0$ \citep[as was resolved in a range of studies, e.g.][]{Noyes1984ApJ...279..763N,Guedel2004A&ARv..12...71G,Ribas2005ApJ...622..680R,Gondoin2012A&A...546A.117G,Jackson2012MNRAS.422.2024J,Shkolnik2014AJ....148...64S,Matt2015ApJ...799L..23M,Johnstone2015A&A...577A..28J,Johnstone2021A&A...649A..96J,Nunez2016ApJ...830...44N,Eggenberger2019A&A...626L...1E,McDonald2019ApJ...876...22M,Magaudda2020A&A...638A..20M}. As different stars are not born with the same rotation rates, those that rotate faster remain in the saturation regime longer compared to slow rotators. This leads to the wide spread in stellar $L_{\rm X}$ (hence, $L_{\rm XUV}$) of stars of the same type at ages before 1-2\,Gyr. This spread is further increased when considering stars of different types: lower mass stars have on average lower luminosities but evolve slower than heavier stars, and are, therefore, characterised by longer \hbindex{saturation times} \citep[e.g.][]{Matt2015ApJ...799L..23M,Johnstone2021A&A...649A..96J}.

\begin{figure}
    \centering
    \includegraphics[width=1\linewidth]{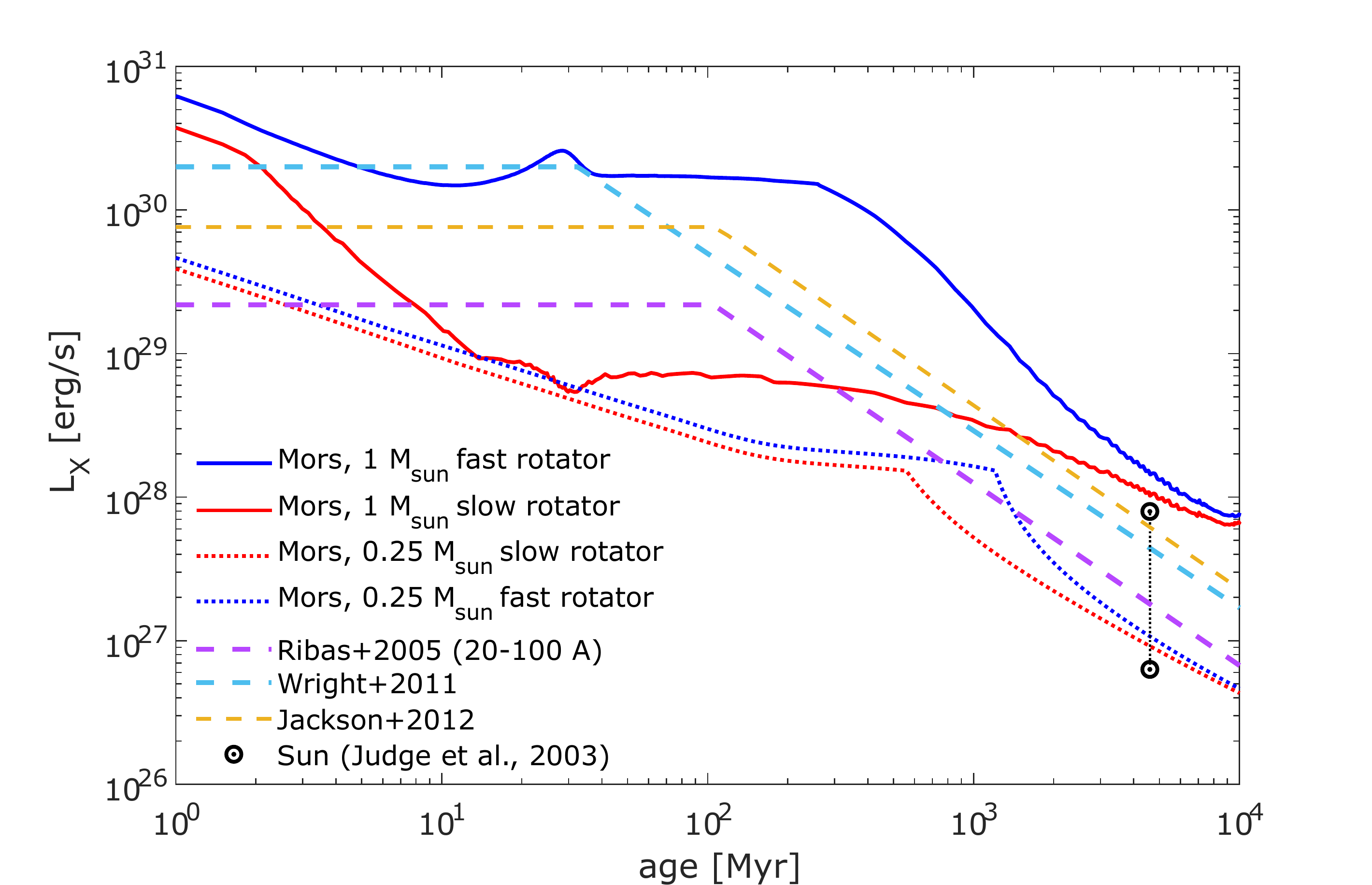}
    \caption{The dependence of stellar X-ray luminosity on time for Sun-like stars (solid and dashed lines) and M-dwarfs (dotted lines) as predicted by a few common models and analytical approximations, as given in the legend: Mors code by \citet{Johnstone2021A&A...649A..96J,Spada2013ApJ...776...87S}; approximation fit on the Sun's evolution by \citet{Ribas2005ApJ...622..680R}; and empirical analytical approximations fit on the observations in young open clusters by \citet{Wright2011ApJ...743...48W} and \citet{Jackson2012MNRAS.422.2024J}. For the comparison, the plot includes the Sun's $L_{\rm X}$ values during maximum and minimum of solar chromospheric activity \citep{Judge2003ApJ...593..534J}.}
    \label{fig:Lx_age}
\end{figure}

Figure\,\ref{fig:Lx_age} compares the evolution of $L_{\rm X}$ of solar-mass star predicted by a few common analytical approximations \citep{Ribas2005ApJ...622..680R,Wright2011ApJ...743...48W,Jackson2012MNRAS.422.2024J} and the Mors stellar evolution code \citep[][]{Johnstone2021A&A...649A..96J}. The analytical approximations represent empirical fits of stellar rotation periods and X-ray luminosities to the observations in stellar open clusters (hence, providing the average parameters for the star of the given type), while the Mors code relies on the \hbindex{rotation evolution} model by \citet[][with the coefficients fit on stellar observations]{Johnstone2015A&A...577A..28J} and the internal stellar parameters from models by \citet{Spada2013ApJ...776...87S}, which allows considering different possible rotation scenarios. For comparison, Figure\,\ref{fig:Lx_age} includes the $L_{\rm X}(age)$ predicted by the Mors code for the M-dwarf star of 0.25\,$M_{\odot}$ and the $L_{\rm X}$ estimates of the Sun at the maximum and minimum of activity \citep{Judge2003ApJ...593..534J}.

The considered approximations predict the saturation times between $\sim30$ and $\sim105$\,Myr and the $L_{\rm X,sat}$ values between $\sim2\times10^{29}$\,erg/s and $\sim2\times10^{30}$\,erg/s and follow the similar power law after this period. For the Mors code, the saturation times are $\sim2$\,Myr for the slowly rotating star (rotation period of 10 days at the age of 150\,Myr) and $\sim200$\,Myr for the fast rotator (rotation period of 1\,day at 150\,Myr), and $L_{\rm X,sat}$ is about $\sim2\times10^{30}$\,erg/s. During the first gygayear of evolution, the predictions of analytical approximations lie in between the two extremes predicted by the Mors code; at later stages, however, the predictions of the Mors code (fit on a larger sample of stellar observational data compared to each of the approximations) lie somewhat above the other predictions and the solar values. The approximation by \citet[][predicting the lowest $L_{\rm X}$ values among those concerned]{Ribas2005ApJ...622..680R} was fit specifically to reproduce the solar $L_{\rm X}$ values as a proxy star, while the observations indicate that the Sun is quite inactive compared to other stars of similar type.
As can be seen from Figure\,\ref{fig:Lx_age}, during the activity minimum, the solar $L_{\rm X}$ resembles that of \hbindex{M-dwarf stars}.

Overall, the comparison between different models demonstrates that the X-ray (and XUV) values predicted by these models for similar stars can differ by about an order of magnitude at a given time, which can have a crucial effect on the predictions of planetary atmosphere evolution models \citep[see e.g.][]{Pezzotti2021A&A...650A.108P}. This stresses the urge for more thorough X-ray surveys necessary to better constrain the stellar evolution models and hence, the evolution of planetary atmospheres.

On top of the above, the \hbindex{spectral energy distribution} of stars also changes with time. For high XUV fluxes, typical for young stars, the fractions of energy deposited in the X-ray and EUV parts of the spectra are expected to be similar or shifted towards X-ray, while the low XUV fluxes of Gyr-old stars are expected to be EUV-dominated \citep[e.g.][]{Sanz-Forcada2011A&A...532A...6S,Claire2012ApJ...757...95C,Linsky2014ApJ...780...61L,Linsky2020ApJ...902....3L,Johnstone2021A&A...649A..96J}. This has implications on the evolution of planetary atmospheres; as discussed in Section\,`Hydrodynamic escape'
, the atmospheric heating by photons of different energies is not equivalent, and high-energy X-ray photons contribute less to powering the atmospheric mass loss \citep[e.g.][]{Ionov2015SoSyR..49..339I,Guo2016ApJ...818..107G,Odert2020A&A...638A..49O}. Therefore, the evolution of stellar $L_{\rm EUV}$ can be more relevant for planetary atmospheric evolution than that of $L_{\rm X}$. Meanwhile, the direct measurements of EUV of planet-hosting stars are complicated because of the absorption by the interstellar medium and the limited sensitivity of instruments. Besides the solar measurements, the available empirical relations are based on as many as $\sim20$ stars \citep[e.g.][]{Johnstone2021A&A...649A..96J}. This constraint will hopefully improved by upcoming stellar surveys such as The Extreme-ultraviolet Stellar Characterization for Atmospheric
Physics and Evolution \citep[ESCAPE][]{France2022JATIS...8a4006F}.

\section{Atmospheric evolution: from giant planets to super-Earths}

Previous sections summarise the main external factors affecting planetary atmospheres. Now, to model their evolution, one has to account for the changes in the lower atmosphere structure (see the $\sim10^{-1}$--$10^6$\,bar interval in Figure\,\ref{fig:PT}) with time. The majority of recent studies \citep[e.g.][]{Baraffe2004A&A...419L..13B,Fortney2010SSRv..152..423F,Lopez2012ApJ...761...59L,Lopez2014ApJ...792....1L,Jin2014ApJ...795...65J,Mordasini2015IJAsB..14..201M,Howe_Burrows2015ApJ...808..150H,Chen2016ApJ...831..180C,kubyshkina2020MNRAS.499...77K,kubyshkina2022MNRAS.510.3039K,Emsenhuber2021A&A...656A..69E,Gu_Chen2023ApJ...953L..27G} tackle this problem by considering the hydrogen-dominated atmosphere on top of the inert solid core \citep[which is typically assumed to be the mixture of silicates and metals but can have a more complex structure including ice/ocean layers; e.g.][]{Valencia2006Icar..181..545V,Dorn2017A&A...597A..37D,Venturini2020A&A...643L...1V}. For the atmospheric part, the adiabatic structure is then solved employing an appropriate \hbindex{equation of state} \citep[e.g.][]{Saumon1995ApJS...99..713S,Haldemann2020A&A...643A.105H}, where the boundary conditions are set by the stellar irradiation and atmospheric mass loss model at the upper boundary and the dissipation of the \hbindex{post-formation luminosity} at the lower boundary.

The thermal evolution of a planet can be constrained by integrating the \hbindex{energy balance equation} with time \citep{Nettelmann2011ApJ...733....2N}. Then, the heat change of each envelope mass shell can be estimated as
\begin{eqnarray}\label{eq:T_evol}
    \int_{M_{\rm core}}^{M_{\rm pl}}dm\frac{T(m,t)dS}{dt} &=& \nonumber\\  
    = - 4\pi R_{\rm pl}(t)^{2}\sigma_{\rm SB}(T_{\rm int}(t)^{4} + (1-A)T_{\rm eq}(t)^4) &+& 
    L_{\rm radio}(t) - c_{\rm v}M_{\rm core}\frac{dT_{\rm core}(t)}{dt}\,,
\end{eqnarray}
where the first term on the right side represents the total internal luminosity (the sum of the own core luminosity and the reflected bolometric irradiation from the star), $L_{\rm radio}$ represents the extra heating of the core due to the decay of the radioactive elements, and the last term describes the cooling of the core.
$M_{\rm core}$ and $T_{\rm core}$ are the mass and the temperature of the planetary core, and $c_{\rm v}$ is its heat capacity.
Along with $L_{\rm radio}$, one can consider some additional heating sources, such as the bloating luminosity in the case of inflated hot Jupiters \citep[which can be caused by the wide range of different physical mechanisms; e.g.][]{Fortney2021JGRE..12606629F} or the heating due to the deuterium burning for the brown dwarfs.

\begin{figure}
    \centering
    \includegraphics[width=1\linewidth]{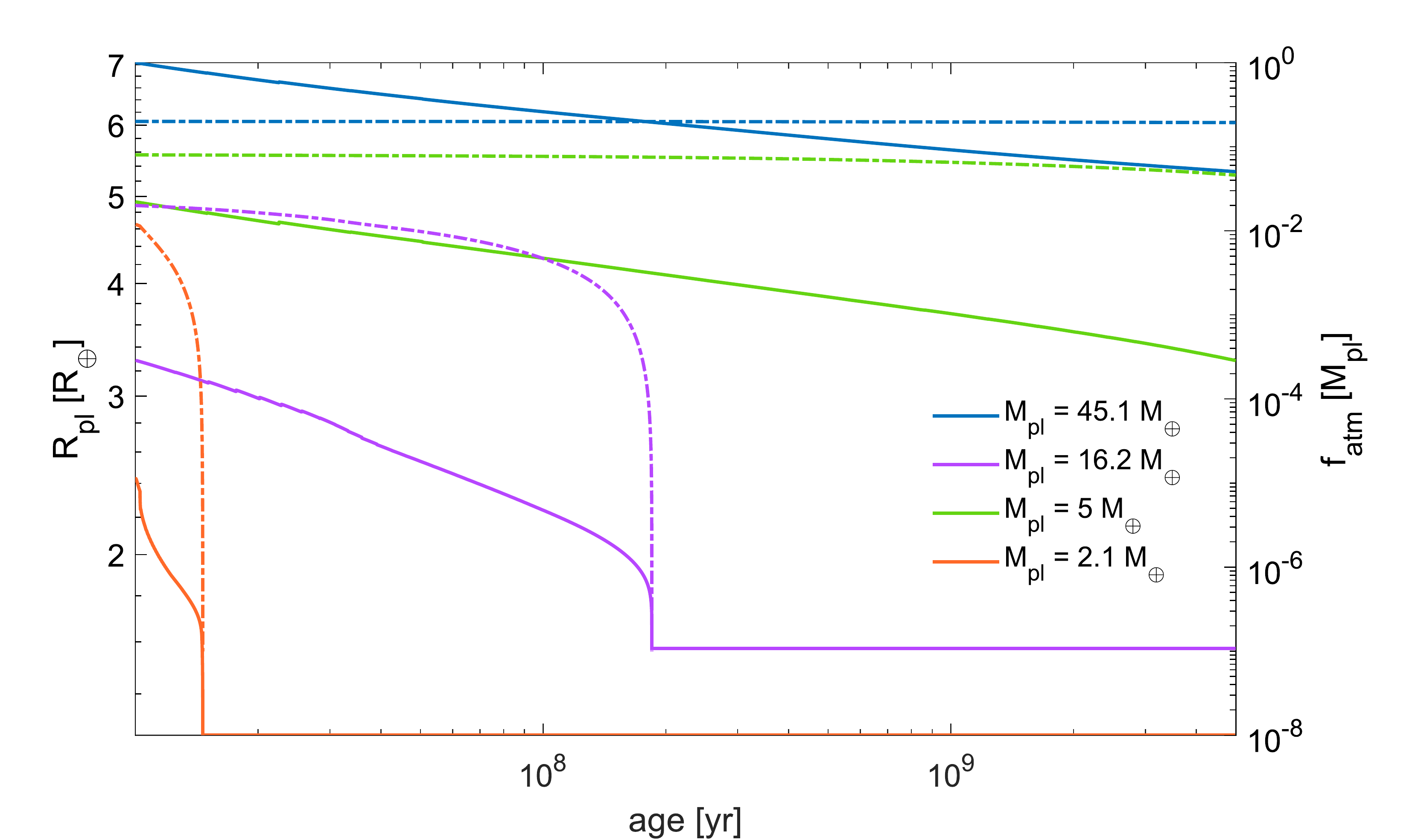}
    \caption{Evolution of planetary radii (solid lines and the left Y-axis) and atmospheric mass fractions (dash-dotted lines and the right Y-axis) with time for planets of different masses (colour-coded in the legend) evolving at $\sim0.0909$\,AU orbit around a solar mass star, as predicted by atmospheric evolution models by \citet{kubyshkina2022MNRAS.510.3039K}. The initial atmospheric mass fractions were set according to the approximation by \citet{Mordasini2020A&A...638A..52M}}
    \label{fig:Rpl_fatm_age}
\end{figure}

With the atmospheric mass adjusted by the mass loss processes and the radius of the planet affected by both the mass loss and the thermal evolution of the atmosphere, low to intermediate-mass planets can change their parameters drastically over their lifetime. To illustrate the different possible evolution paths, Figure\,\ref{fig:Rpl_fatm_age} shows the atmospheric evolution of four model close-in planets orbiting at $\sim0.0909$\,AU around a Sun-like host star evolving as a moderate rotator \citep[following][]{Johnstone2021A&A...649A..96J}. 
Thus, these planets evolve in identical conditions, but their masses span from typical super-Earth to sub-Saturn-like planets.

For the {sub-giant planet} (45.1\,$M_{\oplus}$, blue lines), the atmospheric mass remains nearly unaffected throughout the evolution despite strong XUV and bolometric irradiation from the host star ($T_{\rm eq}\sim1000$\,K and $F_{\rm XUV}\sim$1800--76600\,erg/s/cm$^2$) due to its high gravitational potential. The atmospheric escape rates vary from $\sim1.7\times10^{11}$\,g/s at the beginning of the evolution to $\sim5.7\times10^{9}$\,g/s at the age of 5\,Gyr and the planet loses in total about 3\% of its initial atmosphere (or about 0.3\,$M_{\oplus}$). Along the whole evolution path, the atmospheric mass loss occurs in the XUV-driven regime, with the loss rates almost linearly correlated with stellar EUV flux. 
The changes in planetary radius, therefore, are mainly controlled by the planet's cooling and contraction.
Such an evolutionary path is typical for close-in sub-giant planets, though it was shown that under more extreme conditions the Saturn-like planets can lose a considerable fraction of their atmospheres through the atmospheric mass loss \citep[e.g.][]{Pezzotti2021A&A...654L...5P,Hallatt2022ApJ...924....9H,Kubyshkina_Fossati2022A&A...668A.178K}.

For even more massive \hbindex{hot-Jupiter} type planets with voluminous atmospheres (comprising most of planetary mass), the mass loss plays an even lesser role in the atmospheric evolution; these planets tend to lose no more than about 1\% of their primordial atmospheres throughout the evolution, even though at very close-in orbits their atmospheric mass loss rates can be as high as $10^{12}$--$10^{13}$\,g/s \citep[e.g.][]{Koskinen2022ApJ...929...52K,Huang2023ApJ...951..123H}. Due to the self-gravity of massive atmospheres, the structural models typically predict that the radii of such planets would not significantly exceed that of Jupiter. However, the atmospheric evolution of hot giant planets can differ from the case of the sub-Saturn-like planet considered in Figure\,\ref{fig:Rpl_fatm_age}, as besides the relaxation of primordial luminosity (cooling) these planets can be subject to additional heating in the lower atmospheric layers which suppresses the planet's contraction \citep[the effect known as the \hbindex{radius inflation}, e.g.][]{Bodenheimer2001ApJ...548..466B,Baraffe2003A&A...402..701B,Laughlin2011ApJ...729L...7L,Demory2011ApJS..197...12D,Fortney2021JGRE..12606629F}. This heating can be attributed to various physical processes including tidal interactions with host star \citep[e.g.][]{Bodenheimer2001ApJ...548..466B,Arras2010ApJ...714....1A}, hydrodynamic heat transport toward the planetary interior \citep[e.g.][]{Showman2002A&A...385..166S,Youdin2010ApJ...721.1113Y,Tremblin2017ApJ...841...30T,Sainsbury-Martinez2019A&A...632A.114S} and heating due to induced currents deep in the atmosphere \citep[Ohmic dissipation or induction heating;][]{Batygin2010ApJ...714L.238B,Perna2010ApJ...724..313P,Wu_Lithwick2013ApJ...763...13W,Ginzburg2016ApJ...819..116G,Kislyakova2018ApJ...858..105K,Knierim2022A&A...658L...7K}. More details can be found in chapter `Mass-Radius Relations of Giant planets: The Radius Anomaly and Interior Models'.

As the planet's mass decreases towards the Neptune-like range, the atmospheric mass loss starts to play a progressively larger role in planetary evolution. Thus, the 16.2\,$M_{\oplus}$ model planet considered in Figure\,\ref{fig:Rpl_fatm_age} (green lines) loses almost half of its initial envelope, where about a quarter of the initial envelope is lost during the first gygayear of the evolution when the stellar high-energy irradiation is strongest. As in the case of sub-Saturn-like planet, the changes in atmospheric mass loss rates ($\sim1.0\times10^{10}$--$2.6\times10^{11}$\,g/s) are similar to that in stellar EUV flux and the generalised Jeans parameter $\Lambda$ varies between $\sim$30--38, indicating that the atmospheric escape occurs mainly in the XUV-driven regime. However, for planets of similar mass but larger initial atmospheric mass fractions (hence, larger initial radii) or on closer-in orbits, the escape during the first few tens million years can occur in the transition regime between the core-powered and XUV-driven mass loss (see the case of the low-mass planet considered in Section\,`Hydrodynamic escape').

For even lower-mass planets, the primordial hydrogen-dominated atmosphere can be completely removed from the planet by atmospheric escape; such a process is believed to be responsible for the creation of the so-called \hbindex{radius gap} separating the low-density \hbindex{sub-Neptunes} from the dense \hbindex{super-Earths} in radius-orbital period (or radius-irradiation) plane \citep[][]{Fulton2017AJ....154..109F}. The removal of atmospheres can occur on different timescales; Figure\,\ref{fig:Rpl_fatm_age} includes two planets of 2.1\,$M_{\oplus}$ and 5.0$M_{\oplus}$ with atmospheres fully eroded within $\sim$5\,Myr and $\sim$185\,Myr, respectively. The lower-mass planet remains in the transition regime between the core-powered and XUV-driven atmospheric escape for the whole time of its atmospheric lifetime (see the case considered in Section\,`Hydrodynamic escape'); 
the stellar EUV during this period is nearly constant but the mass loss rates drop by about an order of magnitude ($\sim10^{12}$--$10^{11}$\,g/s) due to the fast contraction of the planetary radius. The more massive 5.0\,$M_{\oplus}$ planet shows similar escape patterns, though the mass loss rates in this case are considerably lower ($\sim2.5\times10^{11}$--$1.0\times10^{10}$\,g/s).

The difference in the atmospheric evaporation times between different low-mass planets has no impact on the parameter distribution of Gyr-old planets in radius-period or mass-radius planes. It can, however, have a crucial impact on the formation of secondary atmospheres through outgassing and their stability, particularly for low-mass terrestrial-like planets, as this time defines the thermal state of the planet and stellar input at the time of these atmosphere's formation and early evolution \citep[hence, the parameters at the planetary surface, exobase position and the energy input into thermal and non-thermal escape processes; see e.g.][]{Lammer2018A&ARv..26....2L,Lammer2020SSRv..216...74L}.

\section{Implications for planetary population}

As discussed above, atmospheric evolution is driven by a range of various processes and can span a range of different scenarios depending on the internal and orbital parameters of a planet and the properties of the host star. Thus, for close-in hot and/or low-mass planets, the evolution of their atmospheres is ultimately decided by the mass loss processes, which occur in the hydrodynamic regime and can fully erode the atmosphere of the planet. For very massive and/or cool objects, the atmospheres are weakly affected by escape (which can occur in a Jeans-like regime or be dominated by non-thermal processes) in terms of their mass; the composition of these atmospheres can yet be altered through the escape of specific ions and fractionation of heavier/lighter elements. However, on a major scale, the atmospheres of such planets are likely defined by their primordial properties. More moderate planets fill the gap between these two extreme scenarios.

While for atmospheric compositions we are only starting to get insights from observations, for the more basic parameters, such as planetary masses, radii, and orbital distances, the accumulated data allows for some classification, and a few features were revealed that are commonly explained as a result of atmospheric evolution.
The most studied of them are the so-called \hbindex{radius gap} \citep[i.e. the dearth of planets with radii of $\sim1.8$\,$R_{\rm oplus}$; e.g.][]{Fulton2017AJ....154..109F,Mullally2015ApJS..217...31M,Van_Eylen2018MNRAS.479.4786V,Ho_VanEylen2023MNRAS.519.4056H} and the \hbindex{Neptunian desert} \citep[i.e. the lack of sub-Neptune-like planets at short orbital separations; e.g.][]{Szabo_Kiss2011ApJ...727L..44S,Beauge2013ApJ...763...12B,Lundkvist2016NatCo...711201L,Mazeh2016A&A...589A..75M}. Both of these features are commonly attributed to the work of hydrodynamic escape, removing the primordial atmospheres from planets with masses too low or orbital separations too short to keep them. This explanation is to some extent supported by the geometry of these features (decrease of the largest bare core for the radius valley and closing of the Neptunian desert with increasing orbital period) and its age dependence \citep[e.g.][]{David2021AJ....161..265D,Sandoval2021ApJ...911..117S,Ho_VanEylen2023MNRAS.519.4056H}. However, the exact mechanism of this atmospheric escape process remains debated. Both photoevaporation \citep[XUV-driven escape; e.g.][]{Owen_Wu2016ApJ...817..107O,Jin2018ApJ...853..163J,Sandoval2021ApJ...911..117S} and core-powered mass loss mechanisms \citep[e.g.][]{Ginzburg2018MNRAS.476..759G,gupta_schlichting2019MNRAS.487...24G} can reproduce the general structure of the radius gap, as well as the models combining both effects \citep{Affolter2023A&A...676A.119A}. However, the present-day accuracy of observations and the correlation between the key parameters controlling both types of outflow (XUV flux and temperature of the planet) do not allow for unambiguous identification of the mass loss mechanism contributing at most to shaping the observed population \citep{Rogers2021MNRAS.508.5886R}. Furthermore, alternative approaches exist explaining these features through the formation processes, such as \hbindex{gas-poor formation} \citep[e.g.][]{Stoekl2015A&A...576A..87S,Lee2022ApJ...941..186L}, primordial planetary composition different from the hydrogen-dominated atmospheres on top of rocky cores \citep[e.g.][]{Rogers_Seager2010ApJ...712..974R,Dorn2017A&A...597A..37D,Schiller2018Natur.555..507S,Morbidelli2020A&A...638A...1M,Venturini2020A&A...643L...1V,Lee_Connors2021ApJ...908...32L,Zeng2021ApJ...923..247Z}, \hbindex{planetary migration} \citep[e.g.][]{Izidoro2022ApJ...939L..19I}, and atmospheric erosion by \hbindex{giant impacts} at the early evolution stages \citep[e.g.][]{Bonomo2019NatAs...3..416B,Wyatt2020MNRAS.491..782W}.

The exoplanet population features described above lie in the plane of radius-period diagram, and were mainly diagnosed based on the results of the Kepler mission. The underlying mass distribution of these planets, therefore, remains poorly constrained. However, the increasing number of radial velocity measurements displays a wide spread in density among planets with masses roughly between that of Earth and Saturn \citep[e.g.][]{Weiss2014ApJ...783L...6W,Hatzes2015ApJ...810L..25H,Ulmer-Moll2019A&A...630A.135U}. Consideration in terms of atmospheric mass loss indicates that the mass-radius distribution is likely shaped by the atmospheric escape processes for low-mass planets ($M_{\rm pl}<\sim10-20$\,$M_{\oplus}$) and retains the imprints of primordial parameters for more massive planets; furthermore, a significant fraction of planets with masses of a few tens of $M_{\oplus}$ appear too dense to be explained in frame of the classic atmospheric evolution approach described in this chapter implying the planets starting with substantial hydrogen-helium atmospheres \citep{Kubyshkina_Fossati2022A&A...668A.178K}. These planets could be explained assuming the formation with water-rich composition \citep[e.g.][]{Venturini2020A&A...643L...1V,Zeng2021ApJ...923..247Z} or the gas-poor formation \citep[e.g.][]{Lee2022ApJ...941..186L}.

\section{Conclusions}

This chapter provides a glimpse into the wide range of processes that need to be accounted for to understand the evolution of planetary atmospheres. 
The atmospheric mass loss, which represents the main driver of atmospheric evolution for low- to intermediate-mass planets, can be engaged by different physical mechanisms; its rate can span over a few orders of magnitude depending on planetary mass, extension, and composition of the planetary atmosphere, orbital separation, and parameters of the host star, and none of these parameters is expected to be constant throughout the whole evolution of a planetary system. Some of them remain poorly constrained, including the early evolution of stellar high-energy emission and the primordial parameters of planets.

The increasing number of observations of atmospheric mass loss from giant and Neptune-like planets provides the possibility of deeper insight into the physics of this process and the parameters of planetary upper atmospheres. The uncertainties in planetary and stellar parameter measurements, however, complicate the interpretation of these results; this holds both for the observations of present-day escape from exoplanets and the imprints of this escape population-wise. Therefore, though the general understanding of the mechanisms involved in driving the evolution of planetary atmospheres appears at a good level, a deeper understanding of their relative roles requires further development in the number and quality of observational constraints.

\section{Cross-References}
\begin{itemize}
\item{Characterising Evaporating Atmospheres of Exoplanets}
\item{Upper Atmospheres and Ionospheres of Planets and Satellites}
\item{Stellar Irradiation and Exoplanet Thermal Evolution}
\item{Stellar Coronal Activity and Its Impact on Planets}
\item{Rotation of Planet-Hosting Stars}
\item{Mass-Radius Relations of Giant planets: The Radius Anomaly and Interior Models}
\item{Mass-Radius Relation of Telluric Planets}
\end{itemize}


\bibliographystyle{spbasicHBexo}  
\bibliography{Atmospheres_through_time} 

\end{document}